\documentstyle[11pt,newpasp,twoside]{article}

\def\etal{\hbox{\it et al.}}

\begin{document}

\title{Cosmological Surveys at Submillimetre Wavelengths}

\author{David H. Hughes}

\affil{Instituto Nacional de Astrofisica, Optica y Electronica  (INAOE),
Apartado Postal 51 y 216, 72000, Puebla, Pue., Mexico}







\begin{abstract}
One of the major goals of observational cosmology is to acquire
empirical data that has the diagnostic power to develop the
theoretical modelling of the high-redshift universe, ultimately
leading to an accurate understanding of the processes by which
galaxies and clusters form and subsequently evolve.  New bolometer
arrays operating on the world's largest submillimetre telescopes now
offer a unique view of the high-redshift universe through unbiassed
surveys with unprecedented sensitivity.  For brevity, except when
there is a need to be more specific, the FIR to millimetre wavelength
regime ($\rm 100\mu m < \lambda < 6000\mu m$) will be referred to as
the ``submillimetre'' (submm).  One of the major challenges in this
field is to accurately quantify the star-formation history of
submm-selected galaxies, particularly those at redshifts $> 1$, and
determine their contribution to the submm extragalactic background.
The field of observational cosmology will be revolutionized during the
course of the next 10 years due to the variety of powerful new
ground-based, airborne and satellite facilities, particularly those
operating at FIR to millimetre wavelengths.  This review summarises
the results from the recent blank-field submm surveys, and describes
the future observations that will provide accurate source-counts over
wider ranges of wavelength and flux-density, constrain the spectral
energy distributions of the submm-selected galaxies and accurately
constrain the redshift distribution and submm luminosity function by
removing the current ambiguities in the optical, IR and radio
counterparts.

\end{abstract}

\keywords{galaxy evolution, millimetre, submillimetre, cosmology}

\section{Evidence for Massive Star-Formation at High-Redshift}

In addressing the question `what is the main epoch of metal production
in the universe?', or equivalently, `when did the cosmic star-formation
rate reach its peak value?', a number of separate lines of evidence
suggest that a high-rate of star-formation ($\gg 100
M_{\odot} \rm yr^{-1}$) must have occurred in massive systems at $z \simeq
3$. This evidence includes (i) the demonstration by Renzini (1998),
using clusters of galaxies as probes of the past star-formation and
metal production history, that 30--50\% of the present-day baryons are
currently locked up in massive structures which appear to have formed at
$z > 3$; (ii) the peak in the co-moving number density of AGN (radio
galaxies and quasars) at $z \sim 2$, AGN whose counterparts at
low-redshift are hosted in luminous, massive elliptical galaxies 
($> 2L^{\star}$ - Taylor \etal\ 1996, McClure \etal\ 1999).  At $z \sim
2$ the universe is only 3--4\,Gyrs old, which implies that a sustained
star-formation rate (SFR) $ > 200 M_{\odot} \rm yr^{-1}$ is required to
build a massive elliptical galaxy by $z \sim 2$ (assuming that
galaxies hosting high-z AGN have already converted the major fraction
of their mass into stars); (iii) the recent discovery of elliptical
galaxies at $z \sim 1.5$ which contain stellar populations
with ages of 3--4 Gyrs (Dunlop {\it et al.} 1996, Peacock {\it et al.}
1998).  Regardless of the cosmological model this requires an extreme
formation redshift ($z > 5$) for the initial starburst in these
galaxies; (iv) the dramatic increase in the number of star-forming
galaxies at high-redshift identified in ground-based and HST faint
galaxy samples. Using a Lyman-break colour selection technique
(Steidel \etal\ 1996), more than 3000 galaxies now have photometric
redshifts with $\sim$ 700 galaxies already spectroscopically confirmed
at $z \simeq 2$ (Adelberger priv.comm.), with SFRs $\sim 1-5 \,
h^{-2} M_{\odot} \rm yr^{-1}$.  However the attenuating effects of
dust, inevitably associated with star-formation, means that SFRs
estimated from these rest-frame UV luminosities must be treated as
strict lower-limits. Near-IR observations of rest-frame Balmer-line
emission suggest an upward correction factor to the SFRs of $2-15
\times$ (Pettini \etal\ 1998), whilst more robust measurements of
SFRs, derived from rest-frame FIR luminosities, imply SFRs 600 times
greater than that estimated from the UV luminosity (Hughes {\it et
al.} 1998, Cimatti {\it et al.}  1998); (v) similar evolution seen in
both the radio-source population and the local starburst population,
implying that radio source evolution is a good tracer of the
star-formation history of the Universe, suggests that the SFR density
derived from Lyman-limit galaxies at $z \simeq 3-4$ is under-estimated
by a factor of $\simeq 5$ (Dunlop 1998), and therefore that once again
the star-formation activity in the Universe peaked at $z > 2$; (vi)
the small, but increasing number of submm continuum and 
CO detections of high-$z$ quasars
and radio galaxies, indicate that the host galaxies of these powerful
AGN contain large quantities of metal-enriched molecular gas ($1-10
\times 10^{10} M_{\odot}$, after correcting for gravitational
amplification) which can fuel massive bursts of star-formation 
(Omont \etal\ 1996, Hughes \etal\ 1997,
Combes \etal\ 1999).

Taken together, the observational evidence suggests that much of the
on-going star-formation in the young Universe may be {\em hidden} by
dust from optical surveys and possibly also from IR surveys.  Hence the
{\em transparent} view of the Universe provided by submm observations,
which now have the instrumental sensitivity to detect high-$z$
dust-enshrouded galaxies forming stars at a rate $> 100 M_{\odot}
\rm yr^{-1}$, and the preliminary evidence that galaxies (particularly
massive spheroidal systems) exhibit strong luminosity evolution at
submm wavelengths, demonstrate that comprehensive submm surveys will
provide an important alternative measurement of the star-formation
history of high-$z$ galaxies unhindered by the effects of dust.

\section{Submillimetre Cosmological Surveys}

The star-formation history of the high-$z$ starburst galaxy population
can be determined from an accurate measure of the integral submm
source-counts, the luminosities and redshift distribution of the
submm-selected galaxies. The contribution of the submm sources to the
total FIR--mm background measured by COBE (Hauser \etal\ 1998, Fixen
\etal\ 1998) places an additional strong constraint on the possible
evolution.  By designing a series of cosmological submm surveys,
covering a sufficiently wide range of complementary depths and areas,
it is possible to discriminate between competing models of galaxy
evolution and the epochs of formation of massive galaxies.

The possibility of conducting cosmological surveys at submm
wavelengths has been realised in the last few years with rapid
technological advances in semiconductor materials, wafer fabrication,
filter design and the temperature stability and performance of
cryogenic systems operating at $\sim 100-400$\,mK. This has led to the
development and successful commissioning of sensitive bolometer arrays
({\it e.g.}  SCUBA, SHARC, MAMBO, BOLOCAM), all of which will be
upgraded within the next few years. These cameras operate on largest
telescopes (10-m CSO, 15-m JCMT, 30-m IRAM) and exploit the best
ground-based mm and submm observing conditions.

Despite these recent advances in instrumental sensitivity the primary
reason that submm observations of galaxies at cosmological redshifts
are at all possible is illustrated in Fig\,1. When attempting to
observe galaxies out to extreme redshifts, observational cosmologists
usually suffer the combined effect of cosmological dimming due to the
vast distances, a steeply declining luminosity function and positive
K-corrections. However the steep spectral index of the Rayleigh-Jeans
emission from dust ($\rm S_{\nu} \propto \nu^{2+\beta}, \beta \simeq
1.5$) heated by young massive stars or AGN, which radiates at
30--70\,K and dominates the FIR-submm luminosity, produces a
negative K-correction at submm and mm wavelengths of sufficient
strength to completely compensate for cosmological dimming at
redshifts $z \geq 1$ with the result that, in an Einstein-de Sitter
universe, a dust-enshrouded starburst galaxy of a given luminosity
should be as easy to detect at an extreme redshift $z \simeq 8$ as at
$z \simeq 1$. Note that at millimetre wavelengths sources actually
become brighter with increasing redshift.  The situation is inevitably
less favourable (by a factor of 2-3) for low values of $\Omega$, but
nevertheless this relative ``ease'' of access to the very
high-redshift universe remains unique to submm cosmology
(Blain \& Longair 1993, Blain \& Longair 1996, Hughes \& Dunlop 1998).

By early 2001, the initial extensive programme of extragalactic SCUBA
(850~$\mu$m) surveys conducted on the 15-m JCMT will be completed,
covering areas of 0.002--0.12~deg$^2$ with respective $3\sigma$ depths
in the range $\rm 1.5~mJy < S_{850\mu m} < 8~mJy$.  To ensure these
submm data are fully exploited, the current SCUBA surveys have been
restricted to fields extensively studied at other wavelengths and
which contain deep X-ray, optical, IR and radio imaging data ({\it
e.g.}  Hubble Deep Field, Hawaii Deep Fields, low-$z$ lensing clusters,
CFRS, Lockman Hole, ELAIS).  The results from these first submm
surveys (outlined below) confirm that an era of massive
dust-enshrouded star-formation exists at early epochs ($z > 1$), with
an intensity previously underestimated at optical wavelengths (Smail
\etal\ 1997, Hughes \etal\ 1998, Eales \etal\ 1999, Blain \etal\ 1999,
Barger \etal\ 1999b). However despite the enhancement of observed
sub-mm fluxes of high-$z$ starburst galaxies by factors of 3--10 at
850$\mu$m, the deepest SCUBA surveys todate are still only sensitive
to high-$z$ galaxies with SFRs comparable to the most luminous local
ULIRGs ($\geq 100 M_{\odot} \rm yr^{-1}$).

\begin{figure}[t]
\vspace{9.5cm}
\includegraphics{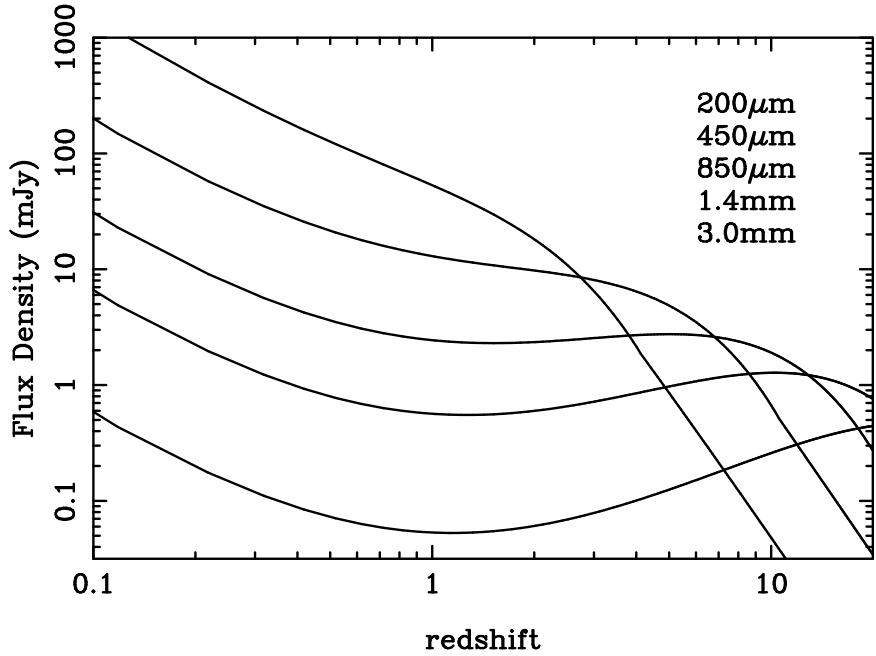}
\caption{\footnotesize Flux density vs. redshift for a galaxy with a
FIR luminosity similar to Arp220, $\rm L_{FIR} \sim 2 \times 10^{12}
L_{\odot}$. At redshift $z = 0.1$, the order of the individual
curves corresponds to same relative order as the wavelength labels. An
$\Omega = 1$ cosmology is assumed.}
\end{figure}

The following preliminary results from the on-going SCUBA surveys have
already made a significant impact on several cosmological questions.

\begin{itemize}

\item
The faint submm source-counts at 850~$\mu$m are reasonably well
determined between 1--10\,mJy ({\it e.g.} Barger \etal\ 1999b) and
significantly exceed a no-evolution model, requiring roughly
$(1+z)^{3}$ luminosity evolution out to $z\sim 1-2$, however a variety
of models are consisitent with the data. The submm background measured
by COBE requires that the SCUBA source-counts must converge at 
$S_{850\mu m} \leq 0.5$\,mJy.

\item 
Submm sources generally appear to be associated with $z > 1$ galaxies,
although it not yet clear whether they necessarily have optical, IR
and radio counterparts.  There is still much debate about the fraction
of submm sources at $z\geq 2$, and the fraction of submm-selected
galaxies that contain an AGN. Currently the most accurate
identifications and spectroscopic redshifts are found in the SCUBA
survey of lensing clusters (Frayer \etal\ 1998, 1999, Ivison \etal\
1998, Barger \etal\ 1999a).

\item
Approximately 30--50\% of the submm background has been resolved into
individual high-$z$ galaxies at flux densities $ S_{850\mu \rm{m}} >
2\,$~mJy, and therefore existing unlensed submm surveys, which are
confusion-limited at about this flux level, are only within a factor
$\sim 4$ in sensitivity of resolving the entire submm/FIR
background. In the SCUBA surveys of highly-lensed clusters it is
possible to measure the source-counts down to a reduced confusion
limit of $S_{850\mu\rm m} \sim 0.5 \rm mJy$ (Blain \etal\ 1998). It now
appears that the majority of submm background is due to a population
of high-$z$ ultraluminous ($L_{\rm FIR} > 10^{12} L_{\odot}$) dusty
galaxies forming stars at rates $ > 100 M_{\odot} \rm yr^{-1}$, {\it
i.e.} similar to the local ULIRG galaxies although the surface density
of the high-$z$ submm population is significantly higher.

\item
At high-redshift ($2 < z < 4$) the submm surveys find $\sim5$ times
the star formation rate observed in the initial optical surveys (Madau
1997). However new optical surveys now agree more closely with the
earlier submm result, finding no significant evidence for a decline in
the star formation density between $z=2$ and $z=4$, when corrections
are applied to the rest-frame UV luminosities to account for
obscuration by dust (Pettini \etal\ 1998, Steidel \etal\ 1998).
\end{itemize}

\section{Limitations on an Understanding of High-z Galaxy Evolution}

Despite the success of the first SCUBA surveys, a number of
deficiencies can be identified in the submm data which prevent a more
accurate understanding of the star-formation history of high-$z$
galaxies.  This review describes these deficiencies and outlines the
future observations which will alleviate the following problems.

\subsection{Constraining the evolutionary models}
To improve the constraints on the competing evolutionary
models provided by the current submm source-counts, it is necessary to
(1) extend the restricted wavelength range of the surveys, (2)
increase the dynamic range of the flux densities over which accurate
source-counts are measured, and (3) increase the number of sources
detected at a given flux level by surveying greater areas.

All these goals can be achieved by conducting future surveys with a
combination of more sensitive, larger format bolometer arrays
operating at 200$\mu$m -- 3\,mm on larger ground-based and airborne
telescopes. Ground-based surveys at mm wavelengths can take advantage
of a more stable and transparent atmosphere which will provide
increased available integration time (to gain deeper survey
sensitivity or greater survey area) and increased flux calibration
accuracy.  Future surveys with more sensitive and larger format arrays
({\it e.g.} BOLOCAM) will allow significantly greater areas to be
covered (hence more sources detected) and will increase the range of the
flux densities over which sources are detected.
Furthermore conducting surveys with larger diameter telescopes ({\it
e.g.} 50-m Large Millimetre Telescope (LMT/GTM), 100-m Green Bank
Telescope) will reduce the beam-size, hence decrease the depth of the
confusion limit (allowing deeper surveys) and improve the positional
accuracy.

\subsection{Future millimetre cosmological surveys}
The SCUBA 850$\mu$m surveys have indicated that the extragalactic
submillimetre background is dominated by a high-$z$ population of
sources with $S_{850\mu m} \leq 1$\,mJy.  At $z \sim 1 - 8$ the same
population of sources will have 1.1mm flux densities in the range $0.3
- 0.6$\,mJy respectively.  The extragalactic confusion limit at
850$\mu$m, estimated from the deepest SCUBA surveys, occurs at a depth
of $\rm 3\sigma < 2 mJy$ and corresponds to a source density N$(S)
\sim 4000 \pm 1500\, \rm deg^{-2}$ at a resolution of 15\,arcsecs.
The ratio of 850$\mu$m (JCMT) and 1.1mm (LMT) beam-areas ($\sim 6''$)
implies that confusion at 1.1\,mm on the LMT will begin to become
significant at a source density of $\rm \sim 24000\, deg^{-2}$.
Extrapolating models that adequately describe the measured 850$\mu$m
source-counts to longer wavelengths suggests that at 1.1\,mm confusion
occurs at $\sim 0.05 - 0.1$\,mJy.  Consequently a deep 1.1\,mm
BOLOCAM/LMT survey has sufficient sensitivity and resolution to detect
the entire submm-mm extragalactic background at a level above the
confusion limit.

The current submm source-counts are measured with varying degrees of
precision at 850$\mu$m between 1--10 mJy (Fig.\,2,), whilst the future
combination of BOLOCAM on the 10-m CSO (2000) and later ($\sim 2002$)
on the 50-m LMT will provide accurate source counts at 1.1\,mm between
0.1 and 100\,mJy.  Table\,1 illustrates the predicted number of
galaxies detected in a series of future 50-hour LMT surveys at 1.1\,mm
with BOLOCAM covering areas of 0.001--70 sq. degrees during the
commissioning phase of the telescope and later, with an improved
surface ($\eta \sim 70\mu$m) during routine operation.  For example in
a similar duration to the original SCUBA HDF survey, a 50-hour LMT
survey at 1.1\,mm (with a $3\sigma$ detection limit of 2\,mJy) will
detect $\sim 2000$ galaxies equivalent to HDF850.1, the brightest
submm source in the HDF ($S_{850\mu\rm{m}} \sim 7$\,mJy).  

\begin{table}[t]
\caption{Number of 5$\sigma$ galaxies detected in alternative 50-hour
BOLOCAM 1.1\,mm surveys during the initial commissioning phase (with a
100$\mu$m r.m.s. surface) of the LMT, and later during routine
operation with an improved surface accuracy (70$\mu$m) and
sensitivity. BOLOCAM sensitivities allow for overheads.}
\begin{center}
\begin{tabular}{rcc|cc}\scriptsize
& \multicolumn{2}{c}{S\tablenotemark{a} = 10 mJy s$^{1/2}$, $\eta\tablenotemark{b} \sim 
100\mu$m} &
\multicolumn{2}{c}{S\tablenotemark{a} =  4 mJy s$^{1/2}$, $\eta\tablenotemark{b} \sim 70\mu$m} \\ \hline
3$\sigma$ survey & Area\tablenotemark{c} & N\,($5\sigma$ galaxies) & 
Area\tablenotemark{c} & N\,(5$\sigma$ galaxies) \\ \hline
0.05 mJy &    0.01  &    6  & 0.06      & 40 \\
0.1 mJy &     0.04  &   15  & 0.25     & 100 \\
0.5 mJy &     1     &  140  & 6    & 860 \\
2.0 mJy &     16    &  400  & 100   & 2460 \\ 
10.0 mJy &   400    &  660  & 2500   & 4120 
\end{tabular}
\end{center}
\vspace{-10mm} 
\tablenotetext{a}{S -- Conservative BOLOCAM Noise
Equivalent Flux Density (NEFD) at 1.1\,mm} 
\tablenotetext{b}{$\eta$ --
Primary aperture r.m.s. surface accuracy ($\mu$m)}
\tablenotetext{c}{Survey area ( $\times 100$ sq. arcmins.)}
\end{table}

\begin{figure*}[t]
\vspace{6.7cm}
\includegraphics{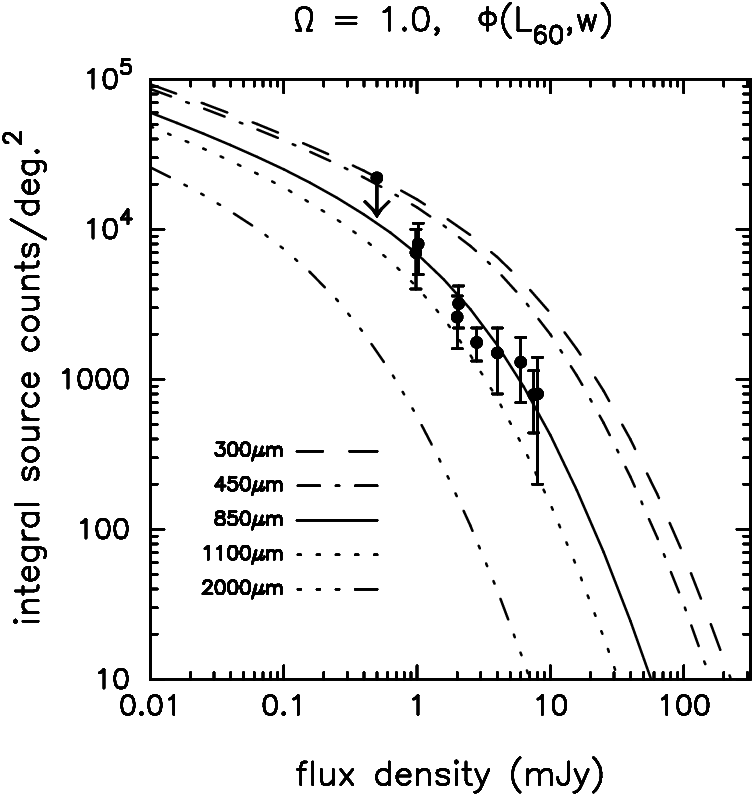}
\includegraphics{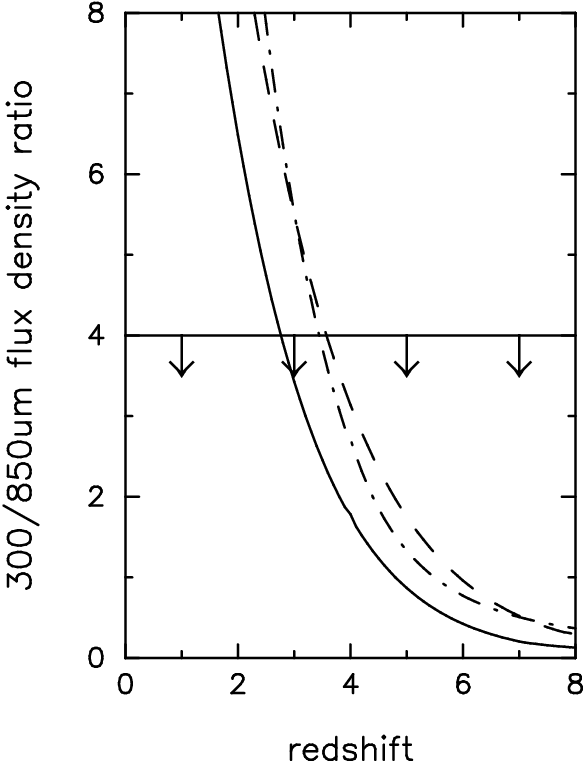}
\caption{\footnotesize The integral number-counts vs. flux
density. The data represent the 850$\mu$m source-counts from the SCUBA
surveys (\S 2). A representative model at
850$\mu$m is extrapolated to derive the expected
number counts at 300--2000$\mu$m. The range of flux densities
represents the extended range that can be surveyed with sufficient
accuracy to fully constrain the competing models.}
\caption{\footnotesize The 300/850$\mu$m flux ratio, appropriate for the
combination of BLAST and SCUBA surveys (\S\,2.1.2), is a powerful
discriminant of redshift. The example of a 5$\sigma$ 850$\mu$m
detection (13~mJy), from the medium-depth UK SCUBA survey, with no
BLAST 300$\mu$m counterpart ($< 50$\,mJy) is indicated by the
horizontal line. The upper limit to the 300/850$\mu$m ratio implies a
redshift $> 3$ assuming the SED of the high-$z$ source is similar to
empirical range of starburst and AGN SEDs represented by the solid
(Arp220), dashed (M82) and dashed-dotted (Mkn231) curves.}
\end{figure*}

\begin{figure*}[t]
\vspace{9.0cm}
\includegraphics{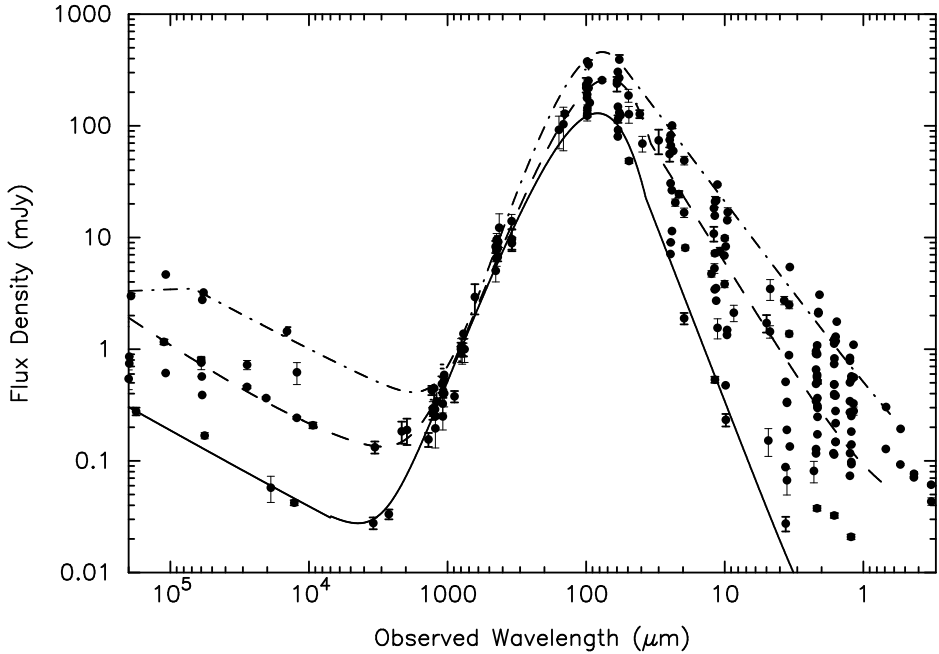}
\caption{\footnotesize 
Radio to IR SEDs of low-z starbursts, quasars, ULIRGS, Seyferts
normalised at 850$\mu$m (Hughes \etal\ 2000, in prep.).  The curves
represent fits to the SEDs of Arp220 (solid), M82 (dashed) and Mkn231
(dashed-dotted). These 3 template SEDs are used to calculate the
range of 300/850$\mu$m colours as a function of redshift in Fig\,3.}
\end{figure*}

\subsection{Ambiguity in the counterparts \& redshifts of submm galaxies}
The current SCUBA surveys (with $15''$ resolution at 850$\mu$m) are
struggling to unambiguously identify the submm sources with their
optical/IR/radio counterparts. Hence the redshift distribution and
luminosities of the submm sources are still uncertain.  This results
directly from the submm positional errors of $\sim 2-3^{\prime\prime}$
that are typical for even the highest S/N submm detections, and from
the lack of submm data measuring the redshifted FIR spectral peak at
200--450~$\mu$m.

The positions of the brightest SCUBA sources ($S_{850\mu m} > 8$\,mJy)
can be improved with mm-interferometric observations. However an IRAM
Plateau de Bure follow-up of the brightest source in the Hubble Deep
Field has demonstrated that even with $\leq 2''$ resolution
and sub-arcsec positional errors, an ambiguous optical identification,
and hence ambiguous redshift remains (Downes {\it et al.} 1999).  It
should be no surprise that submm selected galaxies, including those
with mm-interferometric detections, do not always have optical
counterparts, since high-$z$ galaxies observed in the earliest stages of
formation may be heavily obscured by dust. Indeed this is the most
compelling reason for conducting the submm surveys in the first
instance and therefore searches for the counterparts may be
more successful at near-infrared wavelengths. This was recently
demonstrated by Smail {\it et al.} (1999) who took deep near-IR
($2\mu$m) images of two lensed clusters, previously observed by SCUBA
(Smail {\it et al.} 1997).  The original counterparts were identified
as two bright low-redshift ($z \sim 0.4$) galaxies 5--10 arcsecs
distant from the submm sources.  However the new IR images revealed
two high-$z$ ($z > 2$) IR galaxies, with no optical counterparts,
within 2-3 arcsecs of the SCUBA sources.  The consequence of
these mis-identifications is an inaccurate determination of 
star-formation history of high-$z$ galaxies.  

The uncertainty in the redshift distribution of the submm-selected
galaxies can be significantly reduced by measuring the mid-IR to radio
SEDs of the individual sources.  The power of using mid-IR to radio
flux ratios ({\it e.g.} 15/850$\mu$m, 450/850$\mu$m, 850/1300$\mu$m,
850$\mu$m/1.4~GHz) as a crude measure of the redshift of 
submm-selected galaxies was demonstrated by Hughes {\it et al.} (1998)
during the SCUBA survey of the Hubble Deep Field and has since been
described elsewhere (e.g. Carilli \& Yun 1999, Blain 1999).  The
overall similarity of the IR-radio SEDs of starburst galaxies, ULIRGS
and radio-quiet AGN in the low-$z$ universe
provides a useful {\em template} with which to compare the colours of
high-$z$ submm population (particularly in the absence of information
regarding the relative starburst/AGN contributions). Hence given sufficient
instrumental sensitivity, the FIR--submm--radio colours of a submm source can
discriminate between optical/IR counterparts which are equally
probable on positional grounds alone, but which have significantly
different redshifts, $\delta z \geq 1.5$ (Fig.\,3).

This important technique, and the necessity for sensitive short submm
data (200--500$\mu$m) measuring the rest-frame FIR SEDs of the
individual high-z submm galaxies, without which it remains impossible
to constrain their bolometric luminosities and SFRs, provide the major
scientific justifications behind BLAST, a future ($\sim 2003$) NASA
long-duration balloon-borne large-aperture submm telescope
(P.I. M.\,Devlin, UPenn) operating at an altitude of 140,000 ft.
Table\,2 describes a series of possible BLAST surveys which
demonstrate that even a single 50-hour survey will be able to
follow-up all the wide-area shallow SCUBA surveys observed todate.
For example if there
are 5$\sigma$ SCUBA sources ($S_{850\mu \rm{m}} > 13$\,mJy) with
no {\sl BLAST} 5$\sigma$ counterparts at 300$\mu$m, {\it i.e.} 
$S_{300\mu \rm{m}} < 50$\,mJy, then the 300/850$\mu$m flux ratio
must be $\leq 4$.  This implies that the SCUBA source is most likely a
galaxy at $z \geq 3$ for all typical starburst SEDs (Fig.\,4).  
BLAST will also be an ideal complement to the future BOLOCAM bright mm-surveys
($S_{1.1\,mm} > 4 \rm mJy$) on the CSO and LMT.  A
measurement of the confusion noise due to extragalactic sources at
200--500$\mu$m with a $\sim 2$-m class telescope is an important
secondary goal since the result will influence future FIRST survey
strategies beyond 2007.

\begin{table}[t]
\caption{Predicted number of galaxies detected in possible 50-hour
BLAST surveys. Illustrative redshift distributions are given for
galaxies detected with a S/N $> 5$. A single long-duration balloon
flight will be $\sim 250$ hours duration and will allow several
complementary surveys. The extragalactic confusion limit at 300$\mu$m
will be 20--30\,mJy.}
\begin{center}
\begin{tabular}{cccccc}
\multicolumn{6}{c}{50-hour 300~$\mu$m LDB surveys: D=2.0~m,
NEFD=150 mJy s$^{1/2}$, $\theta = 37''$ } \\  
\hline \hline
survey area         & 
1$\sigma$ depth     & 
\multicolumn{2}{c}{no. of galaxies}   & 
\multicolumn{2}{l}{no. of $>5\sigma$ galaxies} \\ 

(sq. degrees) & 
              & 
$> 5  \sigma$  & 
$> 10 \sigma$  & 
$ z > 1$   & 
$ z > 3$   \\
\hline
0.6  & 5 mJy   & 510 & 160 & 470 & 90 \\
1.2  & 7 mJy   & 600 & 170 & 550 & 90 \\
2.5  & 10 mJy  & 670 & 180 & 600 & 90 \\
5.5  & 15 mJy  & 680 & 150 & 630 & 80 \\
22.2 & 30 mJy  & 600 & 150 & 550 & 65 \\
\end{tabular}
\end{center}
\end{table}

\subsection{Millimetre CO-line spectroscopic redshifts}
An accurate determination of the redshift distribution of
submm-selected galaxies can ultimately be achieved through the
measurement of mm-wavelength CO spectral-line redshifts, without
recourse to having first identified the correct optical or IR
counterparts.  In the high-$z$ Universe the frequency separation of
adjacent mm-wavelength CO transitions is $\rm \delta \nu_{J, J-1} \sim
115/(1+z)$\,GHz. Hence at redshifts $ > 2$, any adjacent pair of CO
transitions are separated by $< 40$\,GHz, similar to the width of the
3\,mm (75--110\,GHz) atmospheric window.
At these frequencies one can expect to detect the most luminous redshifted 
CO transitions from starbursts ($J$=6--5 $\rightarrow$ $J$=3--2).
Therefore, provided one can first pre-select from submm surveys those
galaxies with sufficiently high (but still unknown) redshifts, using
their FIR--mm colours, the availability of a ``CO {\em redshift
machine}'' with a large instantaneous bandwidth ($\Delta\nu \sim
35$\,GHz), operating on large single-dish mm-wavelength telescopes
({\it e.g.} 50-m LMT, 100-m GBT), will offer an incredibly powerful and
more efficient alternative method to determine the accurate redshift
distribution of the submm population.

Whilst future submm surveys will undoubtably detect increasing
numbers of high-$z$ galaxies, an accurate description of their evolutionary
history will not be possible without accurate redshifts and
constraints on their bolometric luminosities (and star-formation
rates). The overall strategy requires the follow-up of submm surveys
with sensitive FIR--submm airborne (SOFIA),  balloon-borne (BLAST) and
satellite observations (SIRTF, FIRST), together with wide-band 
mm-wavelength spectroscopic measurements.\\

\end{document}